\documentstyle[preprint,aps,prb,epsfig]{revtex}
\begin{document}
\draft
\pagestyle{plain}
\newcommand{\D}{\displaystyle}
\title{\bf Role of anisotropic impurity scattering  
in anisotropic superconductors}
\author{Grzegorz Hara\'n\cite{AA} and A. D. S. Nagi} 
\vspace{0.4cm}
\address{Department of Physics, 
University of Waterloo,  
Waterloo, Ontario, 
Canada, N2L 3G1}
\date{July 9, 1996}
\maketitle

\begin{abstract}
A theory of nonmagnetic impurities in an anisotropic superconductor  
including the effect of anisotropic (momentum-dependent) 
impurity scattering is given. It is shown that for a strongly 
anisotropic scattering the reduction of the pair-breaking effect   
of the impurities is large. For a significant overlap between the  
anisotropy functions of the scattering potential and that of the 
pair potential and for a large amount of anisotropic scattering 
rate in impurity potential the superconductivity 
becomes robust vis a vis impurity concentration. The  
implications of our result for YBCO high-temperature superconductor 
are discussed. The experimental data of electron irradiation-induced $T_c$ 
suppression (Phys. Rev. {\bf B50}, 15967 (1994)) is understood 
quantitatively and a good qualitative agreement with the ion ($Ne^+$) 
damage and $Pr$ substitution-induced $T_c$ decrease data (Phys. Rev. {\bf B50}, 
3266 (1994)) is obtained. 
\end{abstract}

\newpage
There now exists a considerable experimental evidence supporting the 
$d$-wave superconductivity in the cuprates (for review see Refs. 1-4).   
Nevertheless, this scenario still faces some theoretical difficulties. 
One of these is the predicted extreme suppression of the critical temperature 
$T_{c}$ by nonmagnetic impurities.\cite{9,15,16,17,18,20} Experimentally, 
however, the observed suppression of $T_{c}$ by impurities or radiation damage   
in YBCO is much more gradual. \cite{11,21}  
This issue was critically examined by Radtke et al.\cite{15} who considered 
isotropic impurity scattering within the second Born approximation 
by applying the Eliashberg formalism. Their predictions in both weak- and 
strong-coupling theory gave a $T_c$ suppression 
which was close to the \mbox{Abrikosov-Gorkov} scaling function \cite{15,2} 
with an effective impurity scattering rate. This led to an 
approximate universal dependence of  $T_c$ on the planar residual resistivity 
$\rho_0$, which did not depend on the details of the microscopic pairing.  
In order to verify the results of Radtke et al. \cite{15} systematic 
electron irradiation experiments on YBCO were carried out by Giapintzakis 
 et al.\cite{21} The measured initial slope of impurity induced $T_c$ 
suppression was $dT_c/d\rho\sim -0.30K/\mu\Omega cm$, \cite{21} whereas  
the predicted value was in the range from $\sim -0.74K/\mu\Omega cm$ 
to $-1.2K/\mu\Omega cm$. \cite{15} While discussing 
the experimental results in Ref. 12 the authors invoked the issue of 
the anisotropic impurity scattering. 
They understood their data within a model of 
Millis et al. \cite{17} assuming a value of 0.5 for a dimensionless  
parameter $g_I$ which describes the anisotropy of the 
scattering potential and modifies the bare isotropic impurity scattering 
rate $1/\tau$ according to $1/\tau^{\star}=\left(1-g_I\right)/\tau$, 
where $1/\tau^{\star}$ is the effective scattering rate. 
Thus the analysis by Giapintzakis et 
al. \cite{21} brings out the significant role of the anisotropic 
scattering in understanding the impurity effect  on $d$-wave superconductivity  
and calls for more detailed theoretical studies.\\ 

In this paper we consider in detail the problem   
of nonmagnetic impurities in an anisotropic superconductor for the case of 
anisotropic (momentum-dependent) impurity scattering by applying weak-coupling 
approximation.  
We find a remarkable change in the $T_{c}$ suppression  
which becomes more gradual when the 
anisotropy function defining anisotropy of the impurity potential  overlaps 
with the anisotropy function of the order parameter. Although our formalism   
is general and valid for any superconducting order parameter described by 
a one dimensional (1D) irreducible representation of the crystal point 
group we discuss the results for a $d$-wave superconductor  
in the context of high temperature superconductivity.  
In a certain limit, 
the effective scattering rate in our model is identical to that of Millis 
et al. \cite{17}  
We compute $T_{c}$ as a function of planar residual 
resistivity. Within a certain range of scattering potential parameters values 
we find a quantitative agreement of our results with the 
electron irradiation data. \cite{21} Also for an appropriate choice of the 
impurity potential coefficients a good qualitative fit to the Pr substitution 
and $Ne^+$-irradiation data \cite{11} is obtained.  
We take $\hbar=k_{B}=1$ throughout the paper.\\ 

We consider randomly distributed nonmagnetic impurities in an anisotropic 
superconductor. Treating the electron-impurity scattering within 
second Born approximation and neglecting the impurity-impurity 
interaction, \cite{2} the normal and anomalous temperature Green's  
functions averaged over the impurity positions read 

\begin{equation}
\label{e2}
G\left(\omega,{\bf k}\right)=
-\frac{i\tilde{\omega}+\xi_{k}}
{\tilde{\omega}^{2}+{\xi_{k}}^{2}+|\tilde{\Delta}\left(
{\bf k}\right)|^{2}}
\end{equation}

\begin{equation}
\label{e3}
F\left(\omega,{\bf k}\right)=
\frac{\tilde{\Delta}\left({\bf k}\right)}
{\tilde{\omega}^{2}+{\xi_{k}}^{2}+|\tilde{\Delta}\left(
{\bf k}\right)|^{2}}
\end{equation}

\noindent
where the renormalized Matsubara frequency 
$\tilde{\omega}\left({\bf k}\right)$ and the renormalized 
order parameter $\tilde{\Delta}\left({\bf k}\right)$ are given by

\begin{equation}
\label{e4}
\tilde{\omega}\left({\bf k}\right)=\omega+i
n_{i}\int |w\left({\bf k}-{\bf k'}\right)|^{2}
G\left(\omega,{\bf k'}\right)\frac{d^{3}k'}{\left(2\pi\right)^{3}}
\end{equation}

\begin{equation}
\label{e4a}
\tilde{\Delta}\left({\bf k}\right)=\Delta\left({\bf k}\right)+
n_{i}\int |w\left({\bf k}-{\bf k'}\right)|^{2}
F\left(\omega,{\bf k'}\right)\frac{d^{3}k'}{\left(2\pi\right)^{3}}
\end{equation}

\noindent
In above $\omega=\pi T(2n+1)$ (T is temperature and n is an integer number), 
$\xi_{k}$ is the quasiparticle energy, $n_{i}$ is impurity (defect) 
concentration, $w\left({\bf k}-{\bf k'}\right)$ is momentum dependent impurity 
potential and $\Delta\left({\bf k}\right)$ is the orbital part of a singlet 
\cite{12} superconducting order parameter defined as  

\begin{equation}
\label{e1}
\Delta\!\left({\bf k}\right)=\Delta e\!\left({\bf k}\right)
\end{equation}

\noindent
where $e\left({\bf k}\right)$ is a real basis function of a 1D 
irreducible representation of an appropriate point group,  
which seems to be good approximation for high $T_{c}$ superconductors. 
\cite{1} We normalize $e\!\left({\bf k}\right)$ by taking 
$\left<e^{2}\right>=1$, 
where $<...>=\int_{FS}dS_k n\left({\bf k}\right)\left(...\right)$ 
denotes the average value over the Fermi surface (FS), 
$n\left({\bf k}\right)$ is the angle resolved FS density 
of states normalized to unity, i.e. $\int_{FS}dS_k n\left({\bf k}\right)=1$,  
and $\int_{FS}dS_{k}$ stands for the integration over the Fermi surface.\\

The impurity scattering potential is assumed to be separable and given by

\begin{equation}
\label{e5}
|w\left({\bf k}-{\bf k'}\right)|^{2}=|w_{0}|^{2}+|w_{1}|^{2}
f\left({\bf k}\right)f\left({\bf k'}\right)
\end{equation}

\noindent
where $|w_{0}|$ ($|w_{1}|$) is isotropic (anisotropic) scattering amplitude 
and $f\left({\bf k}\right)$ is the momentum-dependent anisotropy function.
We assume that the overall scattering rate is determined by the isotropic 
component and impose the constraints

\begin{equation}
\label{e5a}
|w_{1}|^{2}\leq |w_{0}|^{2},\;\;\;\;\;\left<f\right>=0,\;\;\;\;\;
\left<f^{2}\right>=1
\end{equation}

\noindent
Therefore the Fermi surface average of the scattering potential is  
$\left<|w\left({\bf k}-{\bf k'}\right)|^{2}\right>=|w_{0}|^{2}$ and the 
momentum-dependent part in Eq. (\ref{e5}) represents the deviations from the  
isotropic scattering. It is clear that this kind of anisotropic scattering 
cannot affect the properties of the isotropic superconductor, but it can 
play a certain role in the case of a superconductor with an anisotropic 
order parameter.  
Although the structure of scattering potential is postulated in Eq.(\ref{e5}) 
this approach is rather general since no additional assumption about 
$f\left({\bf k}\right)$ is made in contrast to previous methods. \cite{17,6} 
We note from Eq.(\ref{e4}) and from the form of impurity potential  
(Eq. (\ref{e5})) that $\tilde{\omega}$ is ${\bf k}$-dependent.  
This means that the electron self-energy due to impurity scattering  
and consequently the quasiparticle life-time are anisotropic and change  
over the Fermi surface. Further, it yields from Eqs. (\ref{e4a}) and  
(\ref{e5}) that the impurity scattering may change the symmetry of the 
renormalized order parameter $\tilde{\Delta}\left({\bf k}\right)$    
depending on the $f\left({\bf k}\right)$ symmetry.  In this respect our  
approximation differs from that by  Markowitz and Kadanoff \cite{6} who 
assumed only a change of a degree of order parameter anisotropy but not
the anisotropy function itself. Moreover in Ref. 15, the anisotropy of
the order parameter was introduced in a way appropriate for 
weak anisotropy only. In the more recent study of anisotropic scattering 
by Millis et al. \cite{17} the authors also assumed  
that the anisotropic impurity potential does not change the symmetry of 
the electron anomalous self-energy. We may also mention that our approach 
is different than that by Brink and Zuckermann, \cite{7} where the 
scattering potential was essentially isotropic but its amplitude varied  
with the superconducting channels.\\

To proceed further, we restrict the wave vectors of the electron self-energy 
and pairing potential to the Fermi surface and  
replace $\int d^{3}k/\left(2\pi\right)^{3}$ 
by $N_{0}\int_{FS}dS_{k}n\left({\bf k}\right)\int d\xi_{k}$, 
where $N_{0}$ is the overall density of states at the Fermi surface. 
Using Eqs. (\ref{e2}), (\ref{e3}), (\ref{e1}) and (\ref{e5}) 
in Eqs. (\ref{e4}) and (\ref{e4a}) and performing the integration  
over $\xi_{k}$ (particle-hole symmetry of quasiparticle spectrum is assumed) 
we write 

\begin{eqnarray}
\label{e9}
\tilde{\omega}\left({\bf k}\right) & = & 
\omega\left[1+u\left(\omega,{\bf k}\right)\right]\\
\tilde{\Delta}\left({\bf k}\right) & = & \Delta\left[e\left({\bf k}\right)
+e\left(\omega,{\bf k}\right)\right]
\end{eqnarray}
 
\noindent
where $u\left(\omega,{\bf k}\right)$ and $e\left(\omega,{\bf k}\right)$ 
separate into the isotropic (subscript s) and anisotropic (subscript a) 
parts as follows

\begin{eqnarray}
\label{e8}
u\left(\omega,{\bf k}\right) & = & u_{s}\left(\omega\right)
+ u_{a}\left(\omega\right)f\left({\bf k}\right)\\
e\left(\omega,{\bf k}\right) & = & e_{s}\left(\omega\right)
+ e_{a}\left(\omega\right)f\left({\bf k}\right)
\end{eqnarray}

\noindent
which are determined by the self-consistent equations  

\begin{eqnarray}
\label{e10}
u_{s}\left(\omega\right) & = & \Gamma_{0}\int_{FS}dS_{k}n\left({\bf k}\right)
\frac{1+u\left(\omega,{\bf k}\right)}{\left[\tilde{\omega}^{2}+
|\tilde{\Delta}\left({\bf k}\right)|^{2}\right]^{1/2}}
\end{eqnarray}

\begin{eqnarray}
\label{e10a}
u_{a}\left(\omega\right) & = & \Gamma_{1}\int_{FS}dS_{k}n\left({\bf k}\right)
f\left({\bf k}\right)\frac{1+u\left(\omega,{\bf k}\right)}
{\left[\tilde{\omega}^{2}+
|\tilde{\Delta}\left({\bf k}\right)|^{2}\right]^{1/2}}
\end{eqnarray}

\begin{eqnarray}
\label{e10b}
e_{s}\left(\omega\right) &= & \Gamma_{0}\int_{FS}dS_{k}n\left({\bf k}\right)
\frac{e\left({\bf k}\right)+e\left(\omega,{\bf k}\right)}
{\left[\tilde{\omega}^{2}+
|\tilde{\Delta}\left({\bf k}\right)|^{2}\right]^{1/2}}
\end{eqnarray}

\begin{eqnarray}
\label{e10c}
e_{a}\left(\omega\right) &= & \Gamma_{1}\int_{FS}dS_{k}n\left({\bf k}\right)
f\left({\bf k}\right)\frac{e\left({\bf k}\right)+e\left(\omega,{\bf k}
\right)}
{\left[\tilde{\omega}^{2}+
|\tilde{\Delta}\left({\bf k}\right)|^{2}\right]^{1/2}}
\end{eqnarray}

\noindent
In writing above we have introduced the isotropic $\Gamma_{0}$  
and anisotropic $\Gamma_{1}$   
impurity scattering rates ($\Gamma_{1}\leq\Gamma_{0}$) 

\begin{equation}
\label{e10d}
\Gamma_{0}=\pi N_{0}n_{i}|w_{0}|^{2},\;\;\;\;\Gamma_{1}=\pi N_{0}n_{i}|w_{1}|^{2}
\end{equation}

\noindent
The gap function is given by the weak-coupling self-consistent equation 

\begin{equation}
\label{e15}
\Delta\left({\bf k}\right)=-T\sum_{\omega}\sum_{{\bf k'}}
V\left({\bf k}, {\bf k'}\right)
\frac{\tilde{\Delta}\left({\bf k'}\right)}
{\tilde{\omega}^{2}+{\xi_{k'}}^{2}+|\tilde{\Delta}\left(
{\bf k'}\right)|^{2}}
\end{equation}

\noindent
with the phenomenological separable pair potential 
$V\left({\bf k},{\bf k'}\right)$ taken as 

\begin{equation}
\label{e15a}
V\left({\bf k},{\bf k'}\right)=-V_{0}e\left({\bf k}\right)
e\left({\bf k'}\right)
\end{equation}

\noindent
Following standard procedure, \cite{13} we obtain the equation for 
the critical temperature $T_{c}$ as

\begin{equation}
\label{e15b}
\ln\frac{T_{c}}{T_{c_{0}}}=2\pi T_{c}\sum_{\omega>0}
\left[\left(f\left(\omega\right)\right)_{\Delta=0}-\frac{1}{\omega}\right]
\end{equation}

\noindent
with  

\begin{equation}
\label{e15c}
\left(f\left(\omega\right)\right)_{\Delta=0}=
\D\int_{FS}dS_{k}n\left({\bf k}\right)
\frac{e\left({\bf k}\right)}{\tilde{\omega}_{0}\left({\bf k}\right)}
\left[\frac{\tilde{\Delta}\left({\bf k}\right)}
{\Delta}\right]_{\Delta=0}
\end{equation}

\noindent
where $T_{c_{0}}$ is the critical temperature in the absence of impurities and  
\mbox{$\tilde{\omega}_{0}\left({\bf k}\right)=
\tilde{\omega}\left({\bf k}\right)_{\Delta=0}$}.
Using Eqs.(\ref{e9})-(\ref{e10c}), we get for a $\Delta\rightarrow 0$ limit  

\begin{equation}
\label{e11}
\tilde{\omega}_{0}\left({\bf k}\right)=\omega+\Gamma_{0}sign\left(\omega\right)
\end{equation}

\noindent
and 

\begin{equation}
\label{e12}
\left[\frac{\tilde{\Delta}\left({\bf k}\right)}{\Delta}\right]_{\Delta=0}=
e\left({\bf k}\right)+
\frac{|\Gamma_{0}|}{|\omega|}\left<e\right>+
\frac{\Gamma_{1}}{|\omega|+\Gamma_{0}-\Gamma_{1}}
\left<ef\right>f\left({\bf k}\right)
\end{equation}

\noindent
The last two terms on the righthand side of Eq. (\ref{e12}) represent the 
possible impurity induced anisotropy of the renormalized order parameter, which  
is absent in the works of Markowitz and Kadanoff \cite{6} as well as of Millis  
et al. \cite{17}, where $\tilde{\Delta}\left({\bf k}\right)=\tilde{\Delta}
e\left({\bf k}\right)$ is assumed.  
It may be mentioned that $\tilde{\Delta}\left({\bf k}\right)$ and 
$\Delta\left({\bf k}\right)$ have the same anisotropy given by 
$e\left({\bf k}\right)$ function if $f\left({\bf k}\right)=\pm
e\left({\bf k}\right)$ ($\left<e\right>=0$) only. 
Based on Eqs. (\ref{e15c}), (\ref{e11}), and (\ref{e12}) we get from 
Eq. (\ref{e15b})  

\begin{equation}
\label{e16}
\begin{array}{l}
\D\ln\frac{\D T_{c}}{\D T_{c_{0}}}=\left(\left<e\right>^{2}-1\right)
\left(\psi\left(\frac{\D 1}{\D 2}+\frac{\D\Gamma_{0}}{\D 2\pi T_{c}}\right)
-\psi\left(\frac{\D 1}{\D 2}\right)\right)+\\
\\
\D\left<ef\right>^{2}\frac{\D\Gamma_{1}}{\D 2\pi T_{c}}
\D\sum_{n\geq 0} \frac{\D 1}{\D\left(
n+\frac{1}{2}+\frac{\Gamma_{0}}{2\pi T_{c}}\right)}
\frac{\D 1}{\D\left(
n+\frac{1}{2}+\frac{\Gamma_{0}-\Gamma_{1}}{2\pi T_{c}}\right)}
\end{array}
\end{equation}

\noindent
where $\psi\left(z\right)$ is digamma function.\cite{10}  
The first term on the righthand side of Eq. (\ref{e16}) gives 
the $T_{c}$ suppression due to the isotropic scattering. 
Since the second term which couples the anisotropy functions 
$e\left({\bf k}\right)$ and $f\left({\bf k}\right)$ is always 
nonnegative, $T_{c}$ does not decrease as fast 
as for the isotropic scattering only. In other words,  
an anisotropic potential of the form given by Eq. (\ref{e5})  
diminishes the suppression of superconductivity if the 
scalar product $\left<ef\right>$ value is nonzero, which may be 
the case in many cuprate superconducting compounds. For an isotropic 
superconductor $e\left({\bf k}\right)=1$ ($\left<e\right>=1$, 
$\left<ef\right>=0$) and it yields from Eq. (\ref{e16}) that 
the critical temperature does not depend on the impurity  
scattering which is in accordance with the Anderson's theorem. \cite{8}    
Finally, Eq. (\ref{e16}) may be written in a more compact form as 

\begin{equation}
\label{e16a}
\begin{array}{l}
\D\ln\frac{T_{c}}{T_{c_{0}}}=\left(\left<e\right>^{2}+
\left<ef\right>^{2}-1\right)
\left[\psi\left(\frac{1}{2}+\frac{\Gamma_{0}}{2\pi T_{c}}\right)
-\psi\left(\frac{1}{2}\right)\right]+\\
\\
\D\left<ef\right>^{2}\left[\psi\left(\frac{1}{2}\right)-\psi\left(
\frac{1}{2}+\frac{\Gamma_{0}}{2\pi T_{c}}
\left(1-\frac{\Gamma_1}{\Gamma_0}\right)\right)\right]\\
\end{array}
\end{equation}\\

Our model  
has two more dimesionless parameters than the isotropic scattering model.  
First is $\left<ef\right>^2=\left[\int_{FS}dS_{k}n\left({\bf k}\right)
e\left({\bf k}\right)f\left({\bf k}\right)\right]^2$, which describes  
the interplay between  
the pair potential $V\left({\bf k},{\bf k'}\right)$ (Eq. (\ref{e15a}))  
and the anisotropic part of the scattering potential 
$|w\left({\bf k}-{\bf k'}\right)|^{2}$ (Eq. (\ref{e5})). This parameter  
is determined by the symmetry of the superconducting state 
($e\left({\bf k}\right)$) and that of the impurity scattering matrix element 
($f\left({\bf k}\right)$). According to the normalization of the order 
parameter orbital function $\left<e^2\right>=1$ (Eq. (\ref{e1})) and that 
of the anisotropy function of the impurity potential $\left<f^2\right>=1$  
(Eq. (\ref{e5a})) the parameter $\left<ef\right>^2$ takes values  
between 0 and 1. 
When $\left<ef\right>^2=0$ then the $e\left({\bf k}\right)$ and
$f\left({\bf k}\right)$ functions are orthogonal, which means that 
the pair potential $V\left({\bf k},{\bf k'}\right)$ and the impurity 
scattering potential $|w\left({\bf k}-{\bf k'}\right)|^{2}$ do not 
couple and the $T_c$ decrease in Eq. (\ref{e16a}) is due to isotropic 
scattering only. 
On the other hand for $\left<ef\right>^2=1$ we deal with   
$f\left({\bf k}\right)=\pm e\left({\bf k}\right)$ and the pair-breaking 
effect is minimized by the anisotropic part of the scattering potential 
which is proportional to the pair potential. In this case 
Eq. (\ref{e5a}) yields $\left<e\right>=0$ and  Eq. (\ref{e16a}) 
becomes \cite{26}     

\begin{equation}
\label{e16b}
\ln\frac{T_c}{T_{c_{0}}}=
\psi\left(\frac{1}{2}\right)-\psi\left(
\frac{1}{2}+\left(1-g_I\right)
\frac{\Gamma_{0}}{2\pi T_{c}}\right)
\end{equation}

\noindent
with $g_I=\Gamma_1/\Gamma_0$ which leads to the pair-breaking parameter 
$\left(1-g_I\right)\Gamma_0/\left(2\pi T_c\right)$. On the other hand if 
we calculate $g_I$ coefficient defined in Ref. 8 with the impurity potential 
from Eqs. (\ref{e5})-(\ref{e5a}) we get $g_I=\left<e\right>^2+\left<ef\right>^2
\left(\Gamma_1/\Gamma_0\right)$, which reduces to our value of $g_I$ for 
$\left<ef\right>^2=1$ and $\left<e\right>=0$. Thus in this case our pair-
breaking parameter is identical to the one obtained by Millis et al. \cite{17} 
The second parameter in our model ($\Gamma_1/\Gamma_0$) 
represents the amount of anisotropic scattering rate in impurity 
potential normalized by the isotropic scattering rate (Eq. (\ref{e10d})),  
its value ranges also from 0 to 1. For $\Gamma_1/\Gamma_0=0$ we obtain 

\begin{equation}
\label{e17}
\ln\frac{T_{c}}{T_{c_{0}}}=\left(\left<e\right>^{2}-1\right)
\left(\psi\left(\frac{1}{2}+\frac{\Gamma_{0}}{2\pi T_{c}}\right)
-\psi\left(\frac{1}{2}\right)\right)
\end{equation}

\noindent 
which is the isotropic scattering case \cite{9} and yields a considerable 
critical temperature suppression for $\left<e\right>\neq 1$. When 
$\left<e\right>=0$ then Eq. (\ref{e17}) gives a $T_c$ suppression 
curve for a $d$-wave superconductors with isotropic scattering and 
is the weak-coupling version of the form used by Radtke et al. \cite{15} 
In the case of 
strong anisotropic scattering $\Gamma_1/\Gamma_0=1$ and the  $T_{c}$ 
equation reads     

\begin{equation}
\label{e18}
\ln\frac{T_{c}}{T_{c_{0}}}=\left(\left<e\right>^{2}+\left<ef\right>^{2}
-1\right)
\left(\psi\left(\frac{1}{2}+\frac{\Gamma_{0}}{2\pi T_{c}}\right)
-\psi\left(\frac{1}{2}\right)\right)
\end{equation}

\noindent
It is easy to see that the critical temperature suppression becomes 
more gradual now and may be even reversed into $T_{c}$ increase 
for a significant overlap between $e\left({\bf k}\right)$ and 
$f\left({\bf k}\right)$ functions, that is when $\left<ef\right>^2\sim 1$.\\ 

Our results for the dependence of $T_{c}/T_{c_{0}}$ on the
isotropic scattering rate $\Gamma_{0}/2\pi T_{c_{0}}$ are shown in
Figs. 1a-1d for a selection of the model parameters 
$\left<ef\right>^{2}=$ 0.2, 0.4, 0.8 and 0.95 and
$\Gamma_{1}/\Gamma_{0}=$ 0, 0.5, 0.9, 0.95 and 1.0. 
We have assumed here $\left<e\right>=0$. \cite{26}  
Based on these we make the following remarks: (1) In all curves  
the depression of $T_{c}$ in the limit of impurity concentration  
$n_{i}\rightarrow 0$ is given by the initial slope  

\begin{equation}
\label{e19}
\D\frac{d\left(\D\frac{T_c}{T_{c_0}}\right)}
{d\left(\D\frac{\Gamma_0}{2\pi T_{c_0}}\right)}=
-\frac{\pi^2}{2}\left[1-\left<ef\right>^{2}\frac{\Gamma_1}{\Gamma_0}\right]
\end{equation}

\noindent
which decreases drastically as 
\mbox{$\left<ef\right>^{2}\Gamma_{1}/\Gamma_{0}$} approaches unity;
(2) For a given value of $\Gamma_{1}/\Gamma_{0}$, the value of
$\Gamma_{0}/2\pi T_{c_{0}}$ needed to suppress superconductivity
increases as $\left<ef\right>^{2}$ is increased; (3) When there is
a significant overlap between the anisotropy functions
$e\left({\bf k}\right)$ and $f\left({\bf k}\right)$, e.g.
$\left<ef\right>^{2}\sim 0.8$ (Fig. 1c) the value of $\Gamma_{0}/2\pi T_{c_{0}}$
needed to destroy superconductivity is increased considerably
when $\Gamma_{1}/\Gamma_{0}$ becomes large.\\ 

In order to make contact with experiment, we estimate the planar 
residual resistivity $\rho_0$, which is a normal state property and  
according to Eq. (\ref{e11}) depends on the isotropic scattering rate  
$\Gamma_0$ exclusively. It is worth mentionnig here that in the normal  
state the influence of the impurity scattering on the electron  
self-energy is reflected by the frequency rescaling only (Eq. (\ref{e11})),   
and hence the scattering process is characterized by $\Gamma_0$ parameter  
completely. Therefore neither of anisotropic scattering parameters enters  
the equations determining the normal state properties.     
Using a Drude form of the low frequency residual electrical  
conductivity at zero frequency $\sigma=\omega^2_{pl}\tau/4\pi$, 
where $\omega_{pl}$ is the plasma frequency and $1/\tau=2\Gamma_0$, 
we represent \cite{15} the planar residual resistivity in terms of the 
dimensionless pair-breaking parameter $\Gamma_0/2\pi T_{c_0}$    

\begin{equation}
\label{e20}
\D\rho_0\simeq 10.18\times 10^{-2}\times\frac{8\pi^2}{\omega^2_{pl}}
T_{c_0}\left(\frac{\Gamma_0}{2\pi T_{c_0}}\right)\mu\Omega\;cm
\end{equation} 

\noindent
with $\omega_{pl}$ in eV and $T_{c_0}$ in Kelvin. From Eqs. (\ref{e19}) 
and (\ref{e20}) we get the initial slope for a $d$-wave superconductor 

\begin{equation}
\label{e21}
\D\frac{dT_c}{d\rho_0}\simeq -0.615\times \omega^2_{pl}
\left(1-\left<ef\right>^{2}\frac{\Gamma_1}{\Gamma_0}\right)  
K/\mu\Omega\;cm
\end{equation}

\noindent
In the electron irradiation experiment in YBCO Giapintzakis et al. \cite{21} 
obtained $dT_c/d\rho\sim -0.30\pm 0.04\;K/\mu\Omega\;cm$ ($\rho$ is 
resistivity at $145\;K$ and $dT_c/d\rho\simeq dT_c/d\rho_0$).   
Taking the  plasma frequency $\omega_{pl}$ ranging from 1.1 to 
1.4 eV, which is the experimental estimate of $\omega_{pl}$ for YBCO, \cite{21} 
we find from Eq. (\ref{e21}) that the experimental data can be reproduced by 
the anisotropic scattering parameters with values given by a constraint  
$0.55\leq\left(\Gamma_1/\Gamma_0\right)\left<ef\right>^{2}\leq 0.78$.   
The range of values of the scattering parameters stem from an uncertainty 
of the plasma frequency and the $dT_c/d\rho$ measurement accuracy. \cite{21} 
Our calculation focused entirely on a single $CuO_2$ plane 
seems to be a good approximation here since the low-energy electron 
irradiation, used in this experiment,  displaces the oxygen atoms only  
and an appropriate measurement method probes the contribution  
to $T_c$ suppression of only the oxygen defects on the $CuO_2$ planes. \cite{21}  
The two dimensional approach 
is not so justified in the interpretation of the experimental data of Ref.   
11 where $Pr$ substitution and ion ($Ne^+$) damage were applied. 
The $Pr$ substitutes onto the $Y$ site and a similar defect is probably 
induced by ion irradiation since the $T_c$ suppression induced by both 
methods is analogous. Considering this caveat, our theoretical results   
and the experimental data of Ref. 11 are shown in Fig. 2 for an illustrative  
purpose mainly.  
The data were read from Fig. 4 of Sun et al. \cite{11} and the region 
between the curves corresponds to the $T_c$ computed from Eqs. (\ref{e16a}) 
and (\ref{e20}) with $\left<e\right>=0$, $\left<ef\right>^2=0.95$, 
and $\Gamma_1/\Gamma_0=0.96$ for plasma frequencies $\omega_{pl}$ ranging  
from 1.1 to 1.4 eV. We did not try to adjust the amount of anisotropic 
scattering present in our model so as to get a best fit to the data. 
Nevertheless, we note a good qualitative agreement of the 
theoretical results with the experimental data of Ref. 11.  
The experimental data show, however a long tail $T_c$ suppression which 
is not reflected in the computed $T_c$. We think that this feature  
may be due to a slight orthorhombic anisotropy of the system,
\cite{24} which was neglected in our calculation   
by the assumption of $\left<e\right>=0$.\\

Before concluding, we give some critical remarks concerning our 
approach. We have employed a weak-coupling approximation neglecting the 
strong-coupling corrections. We expect that as in Ref. 6, 
the strong-coupling effects would rescale the scattering rates.  
Further, we have neglected the interaction between the nearest $CuO_2$ planes,  
restricting our considerations to a single copper-oxide plane. 
This simplification may not be valid for the 
interpretation of the experimental data of Ref. 11, where the defects 
are not in the $CuO_2$ planes. Finally, we have assumed a model 
separable momentum-dependent impurity potential, which is obviously not  
the most general way of treating the problem, but is more general than 
the one applied in the previous studies. \cite{17,6}\\ 

In summary, we have given a theory of anisotropic impurity scattering 
in anisotropic superconductors. The impurity potential is assumed 
to be separable according to Eq. (\ref{e5}). There are two parameters 
characterizing the scattering anisotropy in our approach. The first of 
them ($\left<ef\right>^2$) represents the interplay between the 
symmetry of the superconducting order parameter and that of the  
impurity potential, the second ($\Gamma_1/\Gamma_0$) gives the amount 
of anisotropic scattering versus the isotropic one. We find that 
for a significant overlap between the pair potential and the 
impurity potential that is for large $\left<ef\right>^2$ values, and 
for a large value of $\Gamma_{1}/\Gamma_{0}$, the anisotropic 
superconductivity becomes robust vis a vis the impurity concentration. 
The experimental data of the electron irradiation-induced $T_c$ 
suppression in YBCO \cite{21} is understood quantitatively within our model. 
We also obtain a good qualitative agreement with the observed $T_c$ 
decrease in YBCO due to ion ($Ne^+$) damage and $Pr$ substitution. \cite{11}\\

We thank J. Taylor for a numerical assistance. This work was  
supported by the Natural Sciences and Engineering Research Council 
of Canada.

\newpage
\section*{Figure Captions}
\noindent
Fig. 1. Normalized critical temperature $T_c/T_{c_{0}}$ 
as a function of the normalized isotropic  
scattering rate $\Gamma_0/2\pi T_{c_{0}}$ for different 
values of the normalized anisotropic scattering rate 
$\Gamma_1/\Gamma_0=$  
0.5 (dotted curve), 0.9 (short-dashed curve), 
0.95 (long-dashed curve), 
1.0 (dot-dashed curve). The solid curve represents the isotropic 
scattering pair-breaking effect ($\Gamma_1/\Gamma_0=0$). We have taken  
$\left<ef\right>^{2}=$ 0.2 (a), 0.4 (b), 0.8 (c), 0.95 (d),  
and $\left<e\right>=0$.\\

\noindent
Fig. 2. Critical temperature $T_{c}$ (area between solid curves) 
of a superconductor with the critical temperature in the absence  
of impurities $T_{c_{0}}=90\;K$ vs residual resistivity for 
$\left<ef\right>^{2}=0.95$, $\Gamma_1/\Gamma_0=0.96$,  
$\left<e\right>=0$ and plasma frequencies $\omega_{pl}$ between 
1.1 and 1.4 eV. The experimental data of Ref. 11 for YBCO are 
shown with circles (Pr substitution) and crosses (ion damage). 

\newpage
\begin{center}
\begin{figure}[p]
\parbox{0.1cm}{\large\vfill $$T_c/T_{c_{0}}$$\vspace{3ex}\vfill }
\parbox{15cm}{\epsfig{file=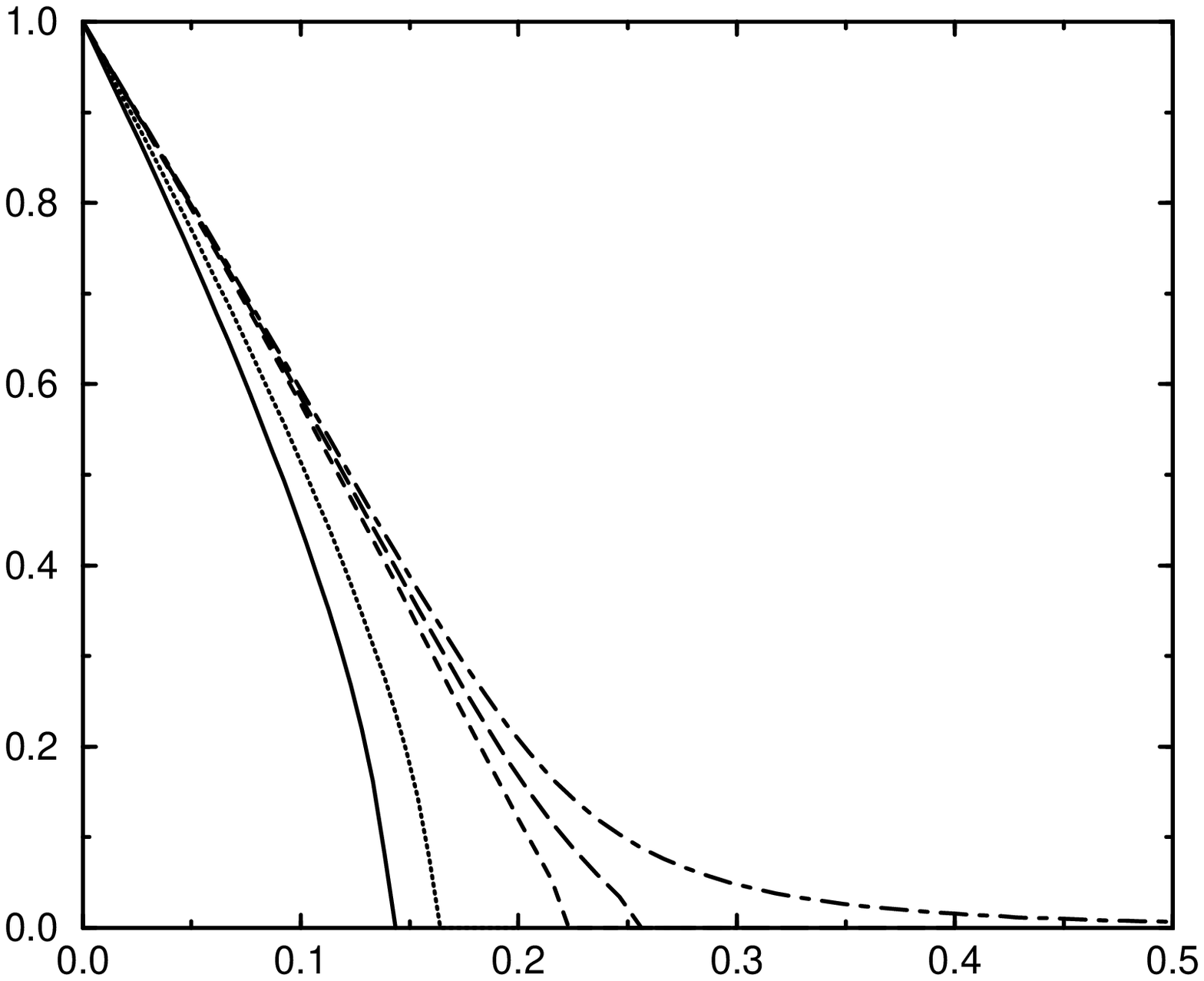,height=15cm,width=15cm} }
\parbox{0.5cm}{\hfill} 
\parbox{18cm}{\large\vspace{-9ex}\hfill $$\Gamma_0/(2\pi T_{c_{0}})\;\;\;\;\;\;$$\hfill }
\end{figure}
\end{center}

\newpage
\begin{center}
\begin{figure}[p]
\parbox{0.1cm}{\large\vfill $$T_c/T_{c_{0}}$$\vspace{3ex}\vfill }
\parbox{15cm}{\epsfig{file=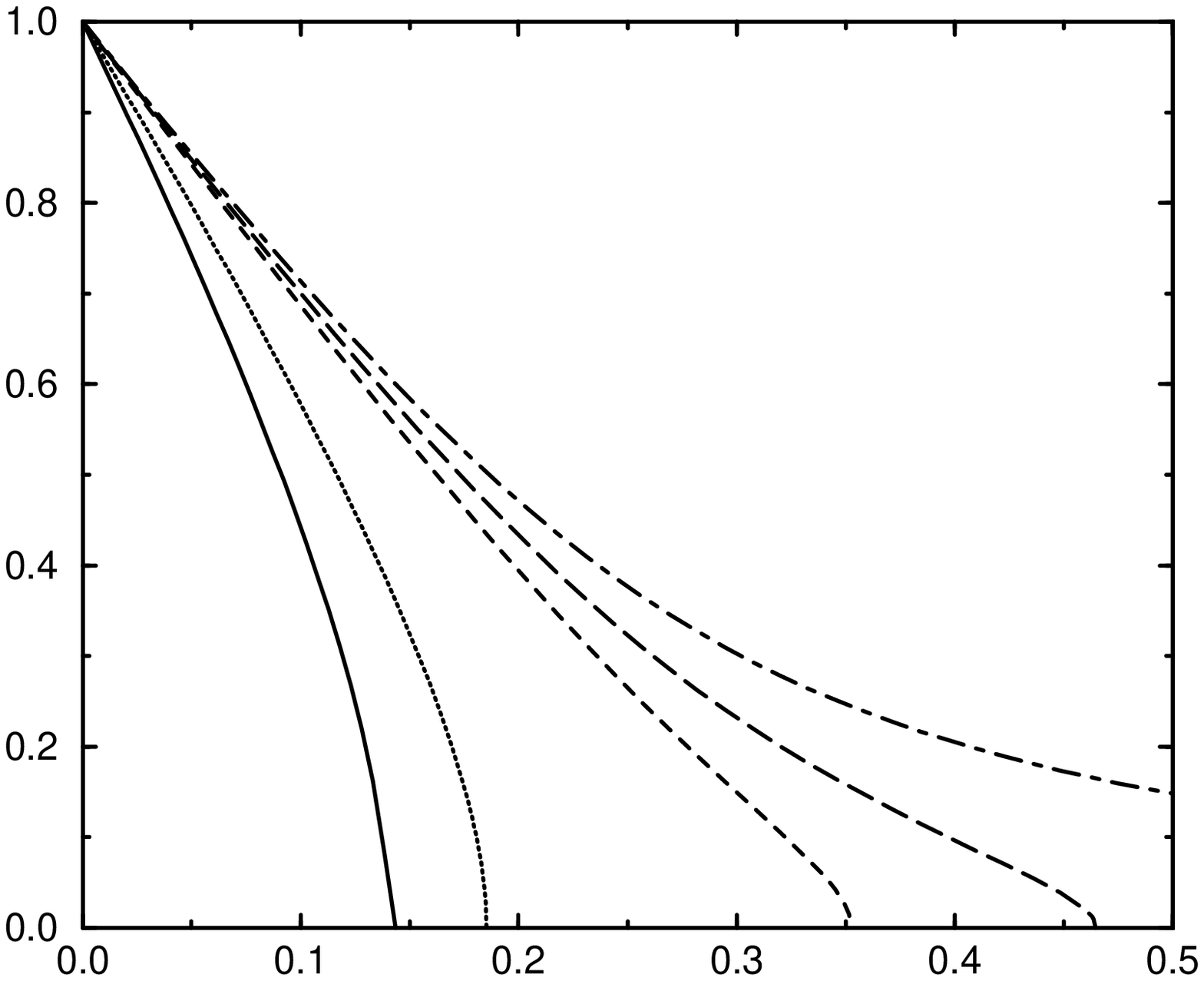,height=15cm,width=15cm} }
\parbox{0.5cm}{\hfill}
\parbox{18cm}{\large\vspace{-9ex}\hfill $$\Gamma_0/(2\pi T_{c_{0}})\;\;\;\;\;\;$$\hfill }
\end{figure}
\end{center}

\newpage
\begin{center}
\begin{figure}[p]
\parbox{0.1cm}{\large\vfill $$T_c/T_{c_{0}}$$\vspace{3ex}\vfill }
\parbox{15cm}{\epsfig{file=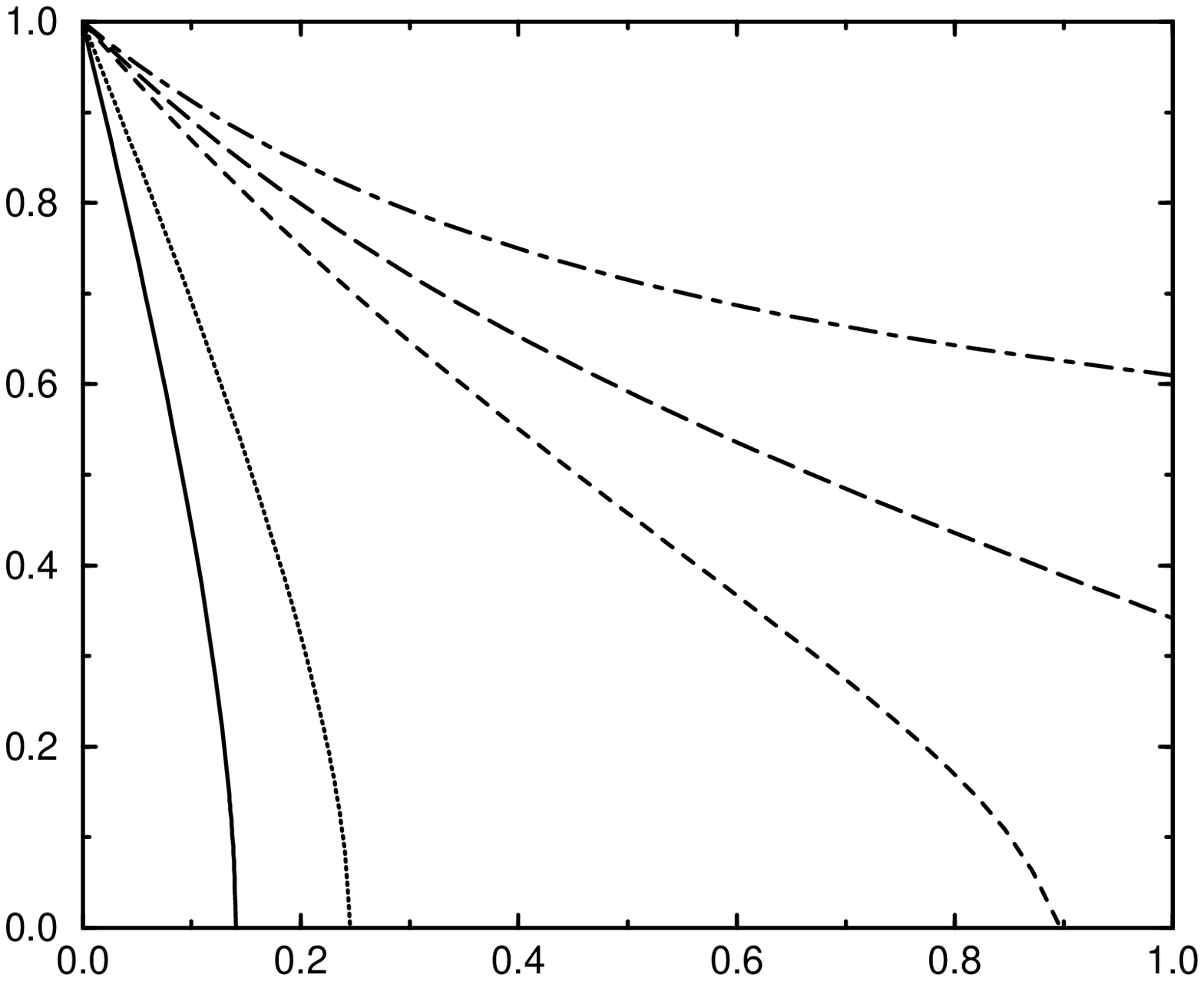,height=15cm,width=15cm} }
\parbox{0.5cm}{\hfill}
\parbox{18cm}{\large\vspace{-9ex}\hfill $$\Gamma_0/(2\pi T_{c_{0}})\;\;\;\;\;\;$$\hfill }
\end{figure}
\end{center}

\newpage
\begin{center}
\begin{figure}[p]
\parbox{0.1cm}{\large\vfill $$T_c/T_{c_{0}}$$\vspace{3ex}\vfill }
\parbox{15cm}{\epsfig{file=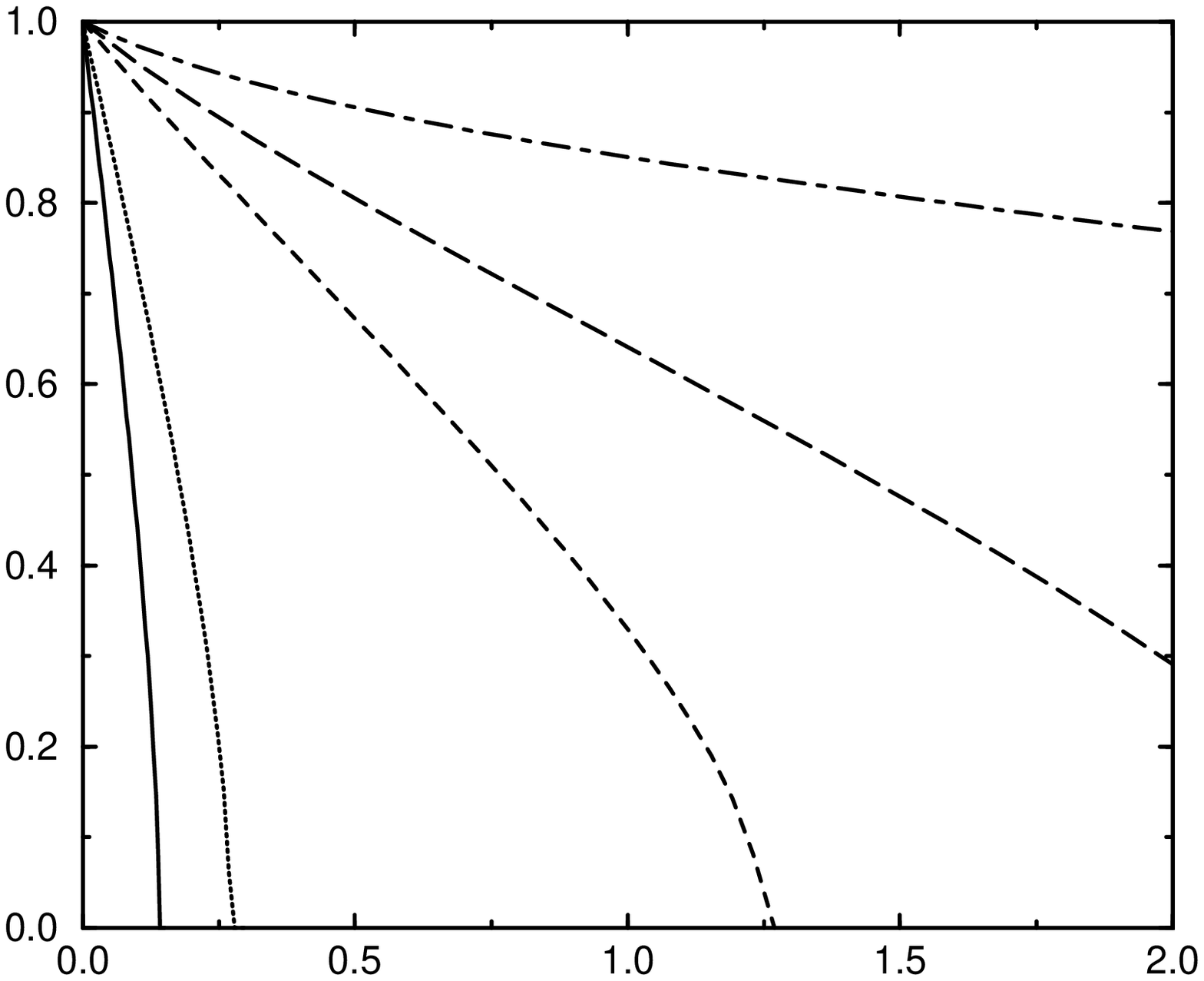,height=15cm,width=15cm} }
\parbox{0.5cm}{\hfill}
\parbox{18cm}{\large\vspace{-9ex}\hfill $$\Gamma_0/(2\pi T_{c_{0}})\;\;\;\;\;\;$$\hfill }
\end{figure}
\end{center}

\newpage
\begin{center}
\begin{figure}[p]
\parbox{0.1cm}{\large\vfill $$T_c\;(K)$$\vspace{3ex}\vfill }
\parbox{15cm}{\epsfig{file=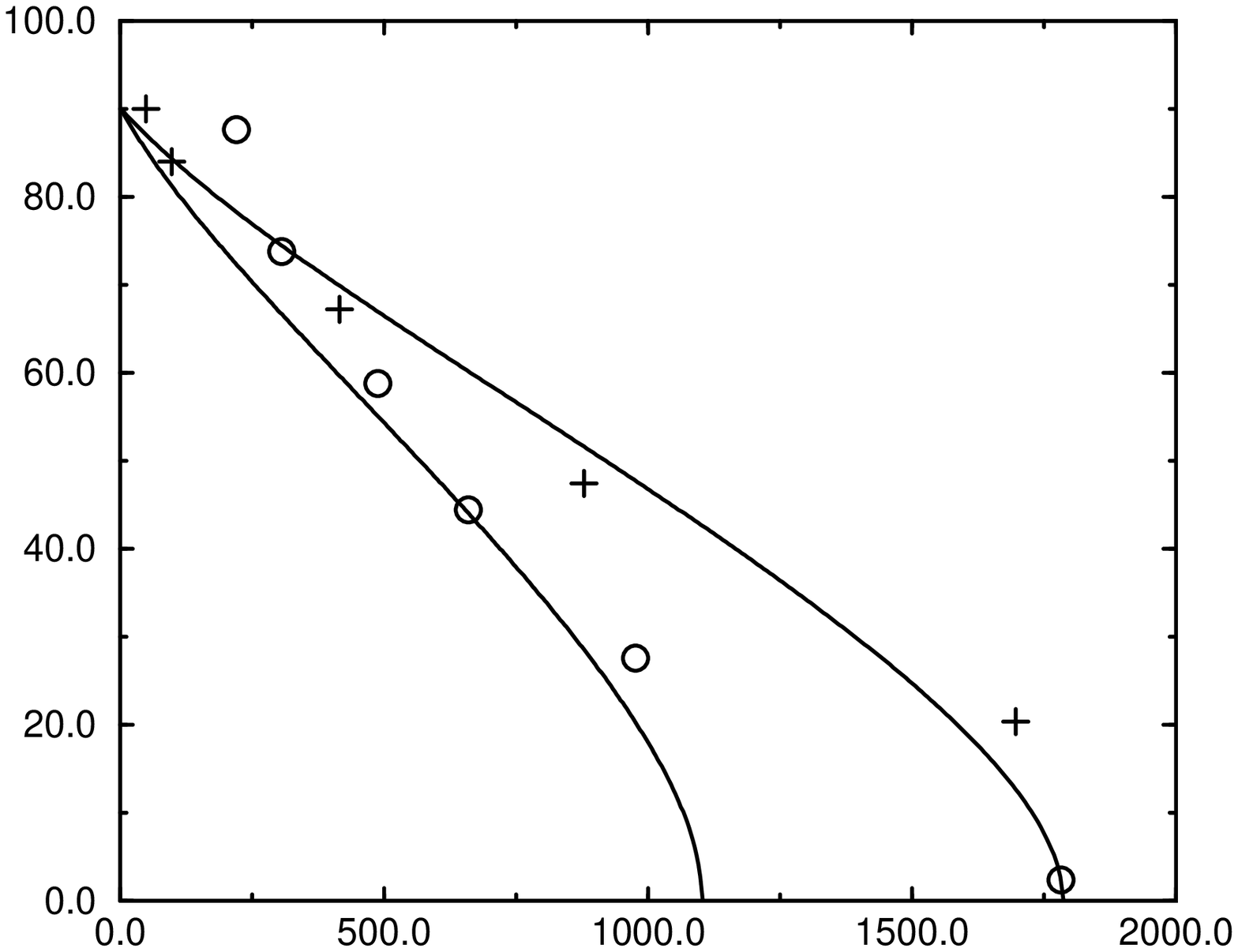,height=15cm,width=15cm} }
\parbox{0.5cm}{\hfill}
\parbox{18cm}{\large\vspace{-9ex}\hfill $$resistivity\:(\mu\Omega cm)$$\hfill}
\end{figure}
\end{center}


\newpage 
\begin{references}
\bibitem[*]{AA} on leave from: Institute of Physics, Politechnika 
Wroc{\l}awska, Wybrze\.ze Wyspia\'nskiego 27, 50-370 Wroc{\l}aw,  
Poland
\bibitem{1} J. Annett, N. Goldenfeld and A. J. Leggett, 
in {\it Physical Properties of High Temperature Superconductors},
Vol. 5, D. M. Ginsberg (ed.), (World Scientific, Singapore, 1996)               
\bibitem{25} D. J. Van Harlingen, Rev. Mod. Phys. {\bf 67}, 515 (1995) 
\bibitem{26} D. Pines and P. Monthoux, J. Phys. Chem. Solids {\bf 56}, 
1651 (1995)
\bibitem{27} D. J. Scalapino, Phys. Reports {\bf 250}, 329 (1995) 
\bibitem{9} A. A. Abrikosov, Physica C {\bf 214}, 107 (1993);
 P. Hohenberg, Sov. Phys. JETP {\bf 18}, 834 (1964)
\bibitem{15} R. J. Radtke, K. Levin, H.-B. Sch\"uttler and M. R. Norman, 
Phys. Rev. {\bf B48}, 653 (1993)
\bibitem{16} T. Hotta, J. Phys. Soc. Japan {\bf 62}, 274 (1993)
\bibitem{17} A. J. Millis, S. Sachdev and C. M. Varma, 
Phys. Rev. {\bf B37}, 4975 (1988)
\bibitem{18} P. Monthoux, A. V. Balatsky and D. Pines, 
Phys. Rev. {\bf B46}, 14803 (1992); P. Monthoux and D. Pines, 
Phys. Rev. {\bf B49}, 4261 (1994)
\bibitem{20} St. Lenck and J. P. Carbotte, Phys. Rev. {\bf B46}, 14850 (1992)
\bibitem{11} A. G. Sun, L. M. Paulius, D. A. Gajewski, M. B. Maple and 
R. C. Dynes, Phys. Rev. {\bf B50}, 3266 (1994); 
J. M. Valles, Jr., A. E. White, K. T. Short, R. C. Dynes, J. P. Garno, 
A. F. Levi, M. Anzlowar and K. Baldwin, Phys. Rev. {\bf B39}, 11599 (1989) 
\bibitem{21} J. Giapintzakis, D. M. Ginsberg, M. A. Kirk and S. Ockers, 
Phys. Rev. {\bf B50}, 15967 (1994)
\bibitem{2} A. A. Abrikosov and L. P. Gorkov, Zh. Eksp. Teor. Fiz. 
{\bf 39}, 1781 (1960) [Sov. Phys. - JETP {\bf 12}, 1243 (1961)]; 
see also   
A. A. Abrikosov, L. P. Gorkov,and I. E. Dzyaloshinski,
{\it Methods of Quantum Field Theory in Statistical Physics}
(Dover, New York, 1975), sec 39
\bibitem{12} Our discussion is also valid for a triplet-paired 
state since the nonmagnetic impurities do not affect 
the spin part of the order parameter.
\bibitem{6} D. Markowitz and L. P. Kadanoff, Phys. Rev. {\bf 131}, 
563 (1963)   
\bibitem{7} D. M. Brink and M. J. Zuckermann, 
Proc. Phys. Soc. {\bf 85}, 329 (1965)
\bibitem{13} K. Maki, in {\it Superconductivity}, R. D. Parks (ed.), 
(Marcel Dekker, New York, 1969), Vol. 2, pp. 1035-1102
\bibitem{10} P. J. Davis, in {\it Handbook of Mathematical Functions},
M. Abramowitz and I. A. Stegun (eds.) (Dover, New York, 1965) 
pp 253-266
\bibitem{8} P. W. Anderson, J. Phys. Chem. Solids {\bf 11}, 26 (1959)
\bibitem{26} $\left<e\right>=0$ for a $d_{x^2-y^2}$ state on a tetragonal 
Fermi surface. Neglecting small orthorhombic distortion of the 
crystal lattice it can be a good approximation for a $d_{x^2-y^2}$-wave 
superconductivity in YBCO. 
\bibitem{24} H. Kim and E. J. Nicol, Phys. Rev. {\bf B52}, 13576 (1995) 
\end{references}
\end{document}